\documentclass[twocolumn,twoside,slac_two]{revtex4}
\usepackage{graphicx}
\usepackage{fancyhdr}
\def\D0bar{\overline D{}^0}
\def\K0bar{\overline K{}^0}
\def\DDbar{D^0 - \overline D{}^0}
\newcommand{\beq}{\begin{equation}}
\newcommand{\eeq}{\end{equation}}
\newcommand{\bey}{\begin{eqnarray}}
\newcommand{\eey}{\end{eqnarray}}
\newcommand{\bdm}{\begin{displaymath}}
\newcommand{\edm}{\end{displaymath}}
\newcommand{\nn}{\nonumber}

\begin{document}

\title{Lifetime difference in $D^0$-${\overline D}^0$ mixing
within R-parity-violating SUSY}

%

\author{Gagik K. Yeghiyan}
\affiliation{Department of Physics and Astronomy, Wayne State University, Detroit, MI 48201, USA}

\begin{abstract}
We re-examine constraints from the evidence for
observation of the lifetime difference in  $D^0$-${\overline D}^0$
mixing on the parameters of supersymmetric models with $R$-parity violation (RPV). We find that RPV SUSY can give large negative contribution to the
lifetime difference. We also discuss the importance of the choice of
weak or mass basis when placing the constraints on RPV-violating
couplings from flavor mixing experiments.
\end{abstract}

\maketitle

\thispagestyle{fancy}


\section{Introduction}
\renewcommand{\theequation}{1.\arabic{equation}}

Meson-antimeson mixing is an important vehicle for indirect search of
New Physics (NP) \cite{48}. $\DDbar$ mixing \cite{36} is the only available
meson-antimeson mixing in the up-quark sector.
The fact that the search is indirect and complimentary to existing constraints
from the bottom-quark sector actually provides parameter space constraints
for a large variety of NP models~\cite{23,6}.

One can write the normalized lifetime difference in $\DDbar$ mixing,
$y_{\rm D} \equiv \Delta \Gamma_{\rm D}/(2\Gamma_{\rm D})$,
as an absorptive part of the $\DDbar$ mixing
matrix~\cite{Petrov:2003un},
\begin{equation}\label{y1}
y_D=\frac{1}{\Gamma_{\rm D}}\sum_n \rho_n
\langle  \overline{D}^0| {\cal H}_w^{\Delta C=1}| n \rangle
\langle n | {\cal H}_w^{\Delta C=1}| D^0 \rangle,
\end{equation}
where $\rho_n$ is a phase space function that corresponds to a
charmless intermediate state $n$.  This relation shows that
$\Delta \Gamma_{\rm D}$ is driven by transitions $D^0, {\overline D}^0 \to n$,
{\it i.e.} physics of the $\Delta C=1$ sector.

It was recently shown~\cite{6} that $\DDbar$ mixing is a rather unique system, where
the lifetime difference can be used to constrain the models of New Physics\footnote{A similar
effect is possible in the bottom-quark sector~\cite{Badin:2007bv}.}. This stems
from the fact that there is a well-defined theoretical limit (the flavor $SU(3)$-limit)
where the SM contribution vanishes and the lifetime difference is dominated
by the NP $\Delta C = 1$ contributions. In real world, flavor $SU(3)$ is, of
course, broken, so the SM contribution is proportional to a (second) power of
$m_s/\Lambda$, which is a rather small number. If the NP contribution to $y_{\rm D}$
is non-zero in the flavor $SU(3)$-limit, it can provide a large contribution
to the mixing amplitude.

To see this, consider a $D^0$ decay amplitude
which includes a small NP contribution, $A[D^0 \to n]=A_n^{\rm (SM)} + A_n^{\rm (NP)}$.
Experimental data for D-meson decays are known to be in a decent agreement
with the SM estimates \cite{47, 28}. Thus, $A_n^{\rm (NP)}$ should be smaller
than (in sum) the current theoretical and experimental
uncertainties in predictions for these decays.

One may rewrite equation~(\ref{y1}) in the form (neglecting the effects of
CP-violation)
\begin{eqnarray}
\nonumber
y_D = \sum_n \frac{\rho_n}{\Gamma_{\rm D}}
A_n^{\rm (SM)} \bar A_n^{\rm (SM)}
+ 2\sum_n \frac{\rho_n}{\Gamma_{\rm D}}
A_n^{\rm (NP)} \bar A_n^{\rm (SM)} + \\
+ \sum_n \frac{\rho_n}{\Gamma_{\rm D}}
A_n^{\rm (NP)} \bar A_n^{\rm (NP)} \ \ .
\label{approx}
\end{eqnarray}
The first term in this equation corresponds to the SM contribution, which vanishes in the
$SU(3)$ limit. In ref. \cite{6}, as well as in the superseding papers \cite{Chen,37},
the last term in (\ref{approx})
has been neglected, thus the NP
contribution to $y_{\rm D}$ comes there solely from the second term, due to
interference of $A_n^{\rm (SM)}$ and $A_n^{\rm (NP)}$.
While this contribution is in general non-zero in the flavor $SU(3)$
limit, in a large class of (popular) models it actually is~\cite{6, 37}. Then,
in this limit, $y_{\rm D}$ is completely dominated by pure $A_n^{\rm (NP)}$
contribution given by the last term in eq.~(\ref{approx})! It is clear that
the last term in equation (\ref{approx}) needs more detailed and careful studies,
at least within some of the NP models.

Indeed, in reality, flavor $SU(3)$ symmetry is broken, so the first term in
Eq.~(\ref{approx}) is not zero. It has been argued~\cite{29} that in fact
the SM $SU(3)$-violating contributions could be at a percent level, dominating the
experimental result, $y_D^{exp} = (0.73 \pm 0.18)\%$ \cite{HFAG}.
The SM predictions of $y_D$, stemming from evaluations of
long-distance hadronic contributions, are rather uncertain.
While this precludes us from placing explicit constraints on
parameters of NP models, it has been argued that, even in this situation,
an upper bound on the NP contributions can be placed~\cite{23} by displaying
the NP contribution only, i.e. as if there were no SM contribution at all.
This procedure is similar to what was traditionally done in the studies of
NP contributions to $K^0-\overline{K}^0$ mixing, so we shall employ it here too.

In order to evaluate importance of the NP contribution, as the flavor $SU(3$) is
broken, counting of suppression powers of $m_s/m_c$ for the SM contribution
versus those of $M_{W}^2/M_{NP}^2$ of the NP contribution must be performed. For the
last term in eq.~(\ref{approx}) to be essential, the following approximate
rule applies: $M_{W}^4/M_{NP}^4 > m_s^2/m_c^2$. This term is of the primary importance
here: the second term in (\ref{approx}) is proven to be $\lesssim 10^{-4}$ in the most
popular SM extensions \cite{6,37,17} and, hence, negligible in general.

The talk is based on the results presented in \cite{basic}.
We revisit the problem of the NP contribution to
$y_{\rm D}$ and provide constraints on R-parity-violating supersymmetric
(SUSY) models as a primary example. It has been recently argued in
\cite{17} that within \slash{\hspace{-0.25cm}R}- SUSY models, new physics
contribution to $y_{\rm D}$ is rather small, mainly because of stringent
constraints on the relevant pair products of RPV coupling constants.
However, this result has been
derived neglecting the transformation of these couplings from the weak
isospin basis to the quark mass basis. This approach seems to be quite
reasonable for the scenarios with the baryonic number violation.
However, in the scenarios with the leptonic number violation,
transformation of the RPV couplings from the weak eigenbasis to the quark
mass eigenbasis turns to be crucial, when applying the existing
phenomenological constraints on these couplings.

We show in that within R-parity-breaking
supersymmetric models with the leptonic number violation, new physics contribution to the
lifetime difference in $\DDbar$ mixing may be large, due to the last term
in eq.~(\ref{approx}). When being large, it is negative
(if neglecting CP-violation), i.e. opposite in sign
to what is implied by the recent experimental evidence for $\DDbar$
mixing.

\section{R-Parity Breaking Interactions: Weak vs Mass Eigenbases}

\setcounter{equation}{0}
\renewcommand{\theequation}{2.\arabic{equation}}
We consider a general low-energy supersymmetric scenario with no
assumptions made
on a SUSY breaking mechanism at the unification scales
$(\sim~(10^{16}~-~10^{18})GeV)$.
The most general Yukawa superpotential for an explicitly broken R-parity
supersymmetric theory is given by
\bey
\nonumber
W_{\slash{\hspace{-0.2cm}R}} = \sum_{i, j, k} \Biggl[
\frac{1}{2} \lambda_{ijk} L_i L_j E^c_k +
\lambda^\prime_{ijk} L_i Q_j D^c_k + \\
+ \frac{1}{2} \lambda^{\prime \prime}_{ijk} U^c_i D^c_j D^c_k
\Biggr] \label{2.1}
\eey
where $L_i$, $Q_j$ are $SU(2)_L$ weak isodoublet lepton and quark
superfields, respectively; $E_i^c$, $U_i^c$, $D_i^c$ are $SU(2$) singlet
charged lepton, up- and down-quark superfields, respectively;
$\lambda_{ijk}$ and $\lambda^\prime_{ijk}$ are lepton number violating
Yukawa couplings, and $\lambda^{\prime \prime}_{ijk}$ is a baryon number
violating Yukawa coupling.
To avoid rapid proton decay, we assume that $\lambda^{\prime \prime}_{ijk} = 0$
and work with a lepton number violating \slash{\hspace{-0.25cm}R}- SUSY model.

For meson-to-antimeson oscillation processes, to the lowest order in the
perturbation theory, only the second term of (\ref{2.1}) is of the importance.
After transforming quark fields from the weak isospin basis
(used in eq.~(\ref{2.1}) to the quark mass eigenbasis, the
relevant R-parity breaking part of the Lagrangian may be presented in
a following form:
\bey
\nn
&& {\cal L}_{\slash{\hspace{-0.2cm}}R} = - \sum_{i, j, k}
\widetilde{\lambda}^\prime_{ijk}
\Big[\widetilde{e}_{i_L} \bar{d}_{k_R} u_{j_L} +
\widetilde{u}_{j_L} \bar{d}_{k_R} e_{i_L} + \\
\nn
&& + \widetilde{d}^{*}_{k_R} \bar{e}_{i_R}^c u_{j_L} \Big]
+ \sum_{i, j, k} \lambda^\prime_{ijk}
\Big[\widetilde{\nu}_{i_L} \bar{d}_{k_R} d_{j_L} + \\
&& \widetilde{d}_{j_L} \bar{d}_{k_R} \nu_{i_L} +
\widetilde{d}^{*}_{k_R}
\bar{\nu}_{i_R}^c d_{j_L} \Big] + h.c. \label{2.10}
\eey
where
\beq
 \widetilde{\lambda}^\prime_{ijk} \ = \
V^*_{jn} \ \lambda^\prime_{ink} \label{2.2}
\eeq
with $V$ being the CKM matrix.

Very often in the literature (see e.g. \cite{6},~\cite{17},~\cite{2}-\cite{5})
one neglects the difference between
$\lambda^\prime$ and $\widetilde{\lambda}^\prime$, based on the fact
that diagonal elements of the CKM matrix dominate over non-diagonal
ones, i.e.
\beq
V_{jn} = \delta_{jn} + O(\lambda) \qquad \mbox{so} \qquad
\widetilde{\lambda}_{ijk} \approx \lambda^\prime_{ijk} +
O(\lambda)
\label{2.11}
\eeq
where $\lambda = \sin\theta_c \sim 0.2$, with $\theta_c$ being the Cabibbo angle.

Notice that relation~(\ref{2.11}) is valid if only there is
{\it no hierarchy} in couplings $\lambda^\prime$. On the other hand, the existing
strong bounds on pair products $\lambda^\prime \times \lambda^\prime$ (or
$\widetilde{\lambda}^\prime \times \widetilde{\lambda^\prime}$)
\cite{1,2,3} and relatively loose bounds on individual couplings
$\lambda^\prime$~\cite{1} suggest that such a hierarchy may exist. We
have shown in the original work \cite{basic} that pair products
$\widetilde{\lambda}^\prime \times \widetilde{\lambda^\prime}$
may be orders of magnitude greater than corresponding products
$\lambda^\prime \times \lambda^\prime$. This fact plays a
crucial role in our analysis.

In what follows, neglecting the transformation of RPV couplings
from the weak eigenbasis to the quark mass eigenbasis would lead to
overestimate of existing phenomenological bounds on these couplings.
As a result, one would get that within R-parity violating supersymmetric
models, NP contribution to the
lifetime difference in $\DDbar$ mixing is rather negligible \cite{17}.
Yet, this result is true if no hierarchy in the values of the relevant
RPV couplings exist. More generally, in presence of such hierarchy, due
to rather loose constraints on the relevant
$\widetilde{\lambda}^\prime \times \widetilde{\lambda^\prime}$
products, RPV SUSY contribution to $y_{\rm D}$ may be of the same order
or even exceed the experimental value.

\section{Dominant Contribution to $y_{\rm D}$}
\setcounter{equation}{0}
\renewcommand{\theequation}{3.\arabic{equation}}
\begin{figure*}[t]
\centering
\includegraphics[width=135mm]{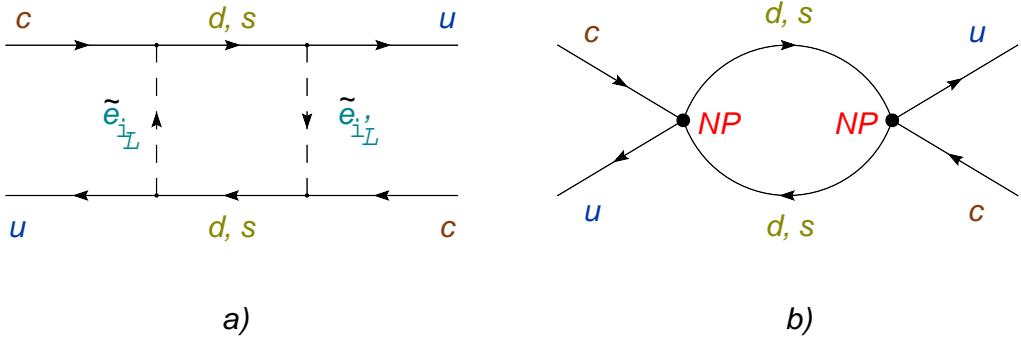}
\caption{Diagrams giving the dominant contribution to $y_{\rm D}$ a) within the
full electroweak theory; b) within the low-energy
effective theory.} \label{f1}
\end{figure*}

Within R-parity violating SUSY models, the dominant contribution to the
lifetime difference in $\DDbar$ mixing comes from the part of $\DDbar$
transition amplitude that occurs when both of $\Delta C = 1$ transitions
are generated by NP interactions, due to exchange of a charged slepton
(see Fig.~\ref{f1}). This contribution to $y_{\rm D}$, denoted here
by $y_{\tilde{\ell} \tilde{\ell}}$, is given by the following formula:
\begin{equation}
y_{\tilde{\ell} \tilde{\ell}} \approx
\frac{- m_c^2
f_D^2 B_D m_D}{288 \pi
\Gamma_D  m_{\tilde{\ell}}^4} \ \Biggl[
\frac{1}{2} + \frac{5}{8} \frac{\bar{B}_D^S}{B_D}
 \Biggr]
\left[\ \lambda_{ss}^2 + \lambda_{dd}^2  \right]
\label{5.3}
\end{equation}
where $f_D$ is D-meson decay constant, $B_D$ and $\bar{B}_D^S$ are
vacuum saturation factors \cite{23} and
\beq
\lambda_{s s} \equiv \sum_{i} \
\widetilde{\lambda}^{\prime *}_{i 1 2} \
\widetilde{\lambda}^{\prime}_{i 2 2},
\  \
\lambda_{d d} \equiv \sum_{i} \
\widetilde{\lambda}^{\prime *}_{i 1 1} \
\widetilde{\lambda}^{\prime}_{i 2 1}
\label{3.8}
\eeq
To simplify the calculations, we assumed
that all the sleptons are nearly degenerate, i.e.
$m_{\tilde{\ell}_i} = m_{\tilde{\ell}}$.

Note that $y_{\tilde{\ell} \tilde{\ell}}$ is non-vanishing in the exact flavor
$SU(3)$ limit. Also, present experimental data still allow
for the slepton masses to be $\sim 100~GeV$ \cite{15}. Finally, present phenomenological
constraints
on the coupling pair products $\lambda_{ss}$ and $\lambda_{dd}$ are rather loose, when
taking into account the transformation of RPV couplings from the weak eigenbasis to
the quark mass eigenbasis (see \cite{basic} for more details). One has
$|\lambda_{ss}| < 0.29$, $|\lambda_{dd}| < 0.29$ or
$\lambda_{ss}^2 < 0.0841$, $\lambda_{dd}^2 < 0.0841$. Thus, as it follows from
our discussion above, $y_{\tilde{\ell} \tilde{\ell}}$ may be quite large.

Indeed, the numerical analysis yields
\beq
-0.12 \left(\frac{100GeV}{m_{\tilde{\ell}}}\right)^4
\leq y_{\tilde{\ell} \tilde{\ell}} < 0 \label{5.12}
\eeq
In other words, $|y_{\tilde{\ell} \tilde{\ell}}|$
may be $\sim 10^{-1}$, if $m_{\tilde{\ell}} = 100$~GeV.

Thus, within R-parity breaking supersymmetric models
with the lepton number violation, new physics contribution
to $D^0 - \bar{D}^0$ lifetime difference is
{\it predominantly negative} and may exceed in absolute
value the experimentally allowed interval. In order to
avoid a contradiction with the experiment
($y_D^{exp} = (0.73 \pm 0.18)\%$ \cite{HFAG}), one must either
have a large positive contribution from the Standard Model, or
place severe restrictions on the values of RPV couplings.
As it follows from \cite{29}, $y_{SM}$ may be as large as
$\sim 1\%$. In what follows, $|y_{new}|$ must be
$\sim 1\%$ or smaller as well. If $|y_{new}| \sim 1\%$,
then, imposing condition
\beq
- 0.01 \leq y_{new} \approx
y_{\tilde{\ell} \tilde{\ell}} \label{5.14}
\eeq
one obtains that either $m_{\tilde{\ell}} > 185$GeV, or
if $m_{\tilde{\ell}} \leq 185$GeV, condition
(\ref{5.14}) implies new bounds on $\lambda_{ss}$ and
$\lambda_{dd}$:
\bey
|\lambda_{ss}| \leq 0.082 \left(\frac{m_{\tilde{\ell}}}{100GeV}
\right)^2 \label{5.15} \\
|\lambda_{dd}| \leq 0.082 \left(\frac{m_{\tilde{\ell}}}{100GeV}
\right)^2 \label{5.16}
\eey
It is interesting to compare the restrictions on $\lambda_{ss}$
and $\lambda_{dd}$, given by (\ref{5.15}), (\ref{5.16}), with
those derived in \cite{23} from study of $D^0 - \bar{D}^0$
mass difference.  Bounds of \cite{23} on $\lambda_{ss}$ and
$\lambda_{dd}$ turn to be about 20 times stronger than our ones.
On the other hand, constraints of ref. \cite{23}
on the
RPV coupling products are derived in the limit when the pure
MSSM contribution to $\Delta m_D$ is negligible. Generally
speaking, the MSSM contribution to $D^0 - \bar{D}^0$ mass
difference is significant even for the squark masses being
about 2TeV. In what follows, the destructive interference of
the pure MSSM and R-parity violating sector contributions
may distort bounds of ref. \cite{23}, making them inessential as
compared to (\ref{5.15}), (\ref{5.16}).

Contrary to this, pure MSSM contributes to $\Delta \Gamma_D$
only in the next-to-leading order via two-loop dipenguin
diagrams. Naturally, this contribution is expected to be small.
In what follows, unlike those of ref. \cite{23},
our constraints on the RPV coupling products
$\lambda_{ss}$ and $\lambda_{dd}$, given by
(\ref{5.15}), (\ref{5.16}), seem to be
insensitive or weakly sensitive to assumptions on the
pure MSSM sector of the theory.

Thus, our main result is that within R-parity breaking
supersymmetric theories with the leptonic number violation,
new physics contribution to $\Delta \Gamma_D$ may be quite
large and is predominantly negative.

\section{Conclusion}

We computed a possible contribution from R-parity-violating
SUSY models to the lifetime difference in $\DDbar$ mixing.
The contribution
from RPV SUSY models with the leptonic number violation
is found to be negative, i.e. opposite in sign
to what is implied by recent experimental evidence, and
possibly quite large, which implies stronger constraints on the
size of relevant RPV couplings.

We discussed currently available constraints on those couplings (especially
on the products of them), available from kaon mixing and rare kaon decays.
We emphasize that the use of these data in charm mixing has to be done
carefully separating the constraints on RPV couplings taken in the mass
and weak eigenbases, given the gauge and CKM structure of $\DDbar$ mixing
amplitudes.

\begin{acknowledgments}

Author is grateful to S. Pakvasa and X. Tata for valuable discussions.

This work has been supported by the grants
NSF~PHY-0547794 and DOE~DE-FGO2-96ER41005.

\end{acknowledgments}

\bigskip 

\end{document}